\pdfoutput=1

\documentclass[final,1p,times]{elsarticle} 
\usepackage{graphicx} 
\usepackage{amssymb} 
\usepackage{amsthm} 
\usepackage{lineno} 

\usepackage{wrapfig}

\journal{Nuclear Physics A} 
\begin{document} 

\begin{frontmatter} 


\title{Anomalous $\phi$ Meson Suppression in Au+Au Collisions at $\sqrt{s_{NN}}=200$~GeV
Measured by the PHENIX Experiment at RHIC}

\author{Maxim Naglis$^{a}$ for the PHENIX collaboration}

\address[a]{Department of Particle Physics, 
 Weizmann Institute of Science, 76100 Rehovot, Israel}

\begin{abstract} 
The PHENIX experiment at RHIC has measured the $\phi$-meson production at mid-rapidity in $p$+$p$,
$d$+Au, Cu+Cu and Au+Au collisions at $\sqrt{s_{NN}}=200$~GeV via the K$^+$K$^-$ decay mode. The
transverse momentum spectra of the $\phi$-meson and the nuclear modification factor 
as a function of centrality are reviewed here.
\end{abstract} 

\end{frontmatter} 



\section{Introduction}
Quarks and gluons are believed to lose energy while traversing the strongly interacting hot
and dense matter produced in relativistic heavy-ion collisions, leading to a significant softening
and broadening of the jets, also known as the phenomenon of jet-quenching. Energy loss of
high-$p_T$ partons manifests itself in the suppression of the particle production at high-$p_T$ in
nucleus-nucleus collisions, compared to expectations from binary scaled $p$+$p$ results.
The amount of suppression and its transverse momentum dependence reflect the opacity of the
medium. The medium-induced effects on particle production are quantified with the nuclear
modification factor: 
\begin{equation}
\label{formula1}
{\rm R}_{AA}(p_T)=\frac{d^2N_{AA}/dydp_T}{\langle N_{coll}\rangle \times d^2N_{pp}/dydp_T},
\end{equation}
where $d^2N_{AA}/dydp_T$ and $d^2N_{pp}/dydp_T$ are the differential yields per event in
nucleus-nucleus and p+p collisions, respectively, and
$\langle N_{coll}\rangle$ is the number of binary nucleon-nucleon collisions averaged over the
impact parameter range of the corresponding centrality.

The studies of hadron production in $p$+$p$, $d$+Au and Au+Au collisions in the first three
runs of RHIC allowed to establish that (i) neutral pions at high-$p_T$ are strongly suppressed in
central Au+Au collisions \cite{pi0_suppress_AuAu}, (ii) no suppression is observed at
high-$p_T$ for direct photons in Au+Au collisions \cite{no_suppress_phot} and (iii) for inclusive
and identified hadrons in $d$+Au collisions \cite{no_suppress_dAu}. Contrary to the strong
suppression of neutral pions, protons show a very different binary-collision scaling behavior
from that of pions in Au+Au collisions \cite{protons_AuAu}. The proton production is enhanced at
intermediate $p_T\approx$~2-5~GeV/$c$ compared to scaled $p$+$p$ results. The 
measurements by the STAR collaboration indicate that at high $p_T$ the nuclear modification   
factor of protons becomes similar to that of pions \cite{protons_star} and also that the
R$_{CP}$ patterns for $K^0_S$ and $K(892)^*$ mesons are different from that of the $\Lambda$ baryon 
(note that $m_{K(892)^*} \approx m_{\Lambda}$) \cite{ks_lambda_star,kstar_lambda_star}. Further
studies of identified hadron production reveal that despite a factor of about 4 difference in
mass, $\eta$ mesons follow the suppression pattern of neutral pions over the entire $p_T$ range of
the measurements \cite{pi0_eta_suppress}. The latter suggests that the differences in the
behavior of baryons and mesons are not related to the difference in their mass but 
rather to the number of constituent quarks. The $\phi$-meson, with a mass comparable to that of
the proton and $\Lambda(1115)$ baryon, but carrying two quarks, differentiates between hadron mass
and number of constituent quark effects. Moreover, being an almost pure $s\bar{s}$ state, it
provides insight on the effects of flavour composition on the hadron suppression pattern. 

\section{Measurements of the $\phi\rightarrow K^+K^-$ with the PHENIX Detector}

The $\phi\rightarrow K^+K^-$ results presented in this contribution were obtained from the data
samples accumulated by the PHENIX experiment \cite{phenix} during  $p$+$p$, $d$+Au, Cu+Cu and
Au+Au collisions at $\sqrt{s_{NN}}=200$~GeV in the 2003-2005 physics runs. 
Charged particle tracking and measurements of
their momentum are accomplished by the high-resolution multi-wire proportional Drift Chambers (DC) and 
the first layer of the Pad Chambers (PC1). The typical value of the momentum resolution is
$\sigma(p_T)/p_T \approx 1.0\%p_T~\oplus~1.1\%$. Charged particle identification (PID) is based on
the particle mass calculated from the momentum and the time-of-flight information derived from the
Time of Flight (TOF) detector or the Lead Scintillator (PbSc) part of the  Electro Magnetic
Calorimeter and the Beam Beam Counters (BBC's). The TOF subsystem with a time resolution 
$\sigma$ $\simeq$ 120~ps and the PbSc with $\sigma$ $\simeq$ 500~ps allow to achieve reliable
pion-kaon separation in the $p_T$ ranges $0.3-2.5$~GeV/$c$ and $0.3-1.0$~GeV/$c$, respectively.
The Zero Degree Calorimeters (ZDC's) and BBC's are dedicated subsystems that determine the
collision vertex and event centrality and also provide the minimum bias (MB) interaction trigger. 
The MB trigger used for $p$+$p$, $d$+Au and Cu+Cu collisions requires a coincidence between the
BBC's with at least one hit in each BBC arm. For Au+Au collisions the MB trigger requires a
coincidence between the BBC's and ZDC's with at least two hits in each BBC arm, and at least one
neutron detected in each ZDC arm. All events used in the analysis are required to have the
collision vertex position along the beam axis within 30 cm of the geometrical center of PHENIX.
There are three different techniques that were used for the $\phi\rightarrow K^+K^-$ mass
reconstruction. The first (``no PID'') does not require identification of charged tracks in the
final state and assumes that all tracks are kaons. The second (``one kaon PID'') requires
identification of only one kaon in the TOF. In the third technique  (``two kaons PID'')
both kaons are identified in the TOF or the PbSc; four different subsystem combinations have been
considered: TOF-TOF, TOF-PbSc$_{EAST}$, PbSc$_{EAST}$-PbSc$_{EAST}$, PbSc$_{WEST}$-PbSc$_{WEST}$.

The $d$+Au data were analyzed using ``no PID'', $p$+$p$ and Cu+Cu using ``no PID'' and ``one kaon
PID'', and Au+Au using ``no PID'' and ``two kaons PID''. 
Combining the results obtained with complimentary techniques the $p_T$ range accessible for the
$\phi$ measurements extends from $\sim$1~GeV/$c$ to 7~GeV/$c$ (to 5.1~GeV/$c$ in $d$+Au).
The three analysis techniques have very different sources of systematic uncertainties and
provide a valuable consistency check.
In every technique all kaon candidates from each event passing the track selection requirements
are combined into unlike-sign pairs. The invariant mass and the transverse momentum for each pair
is then calculated based on the 2-body decay kinematics. The resulting mass distributions contain
both the $\phi$-meson signal and an inherent combinatorial background. The combinatorial
background is estimated either by simultaneously fitting the mass distribution to the sum of a
Breit-Wigner function to account for the signal, convoluted with Gaussian to account for the mass
resolution, and a polynomial function to account for the background, or by an event-mixing
technique described elsewhere \cite{ppg16}. The raw yields are extracted by integrating the
invariant mass distributions in the vicinity 
of the PDG value of the $\phi$-meson mass (1.019~GeV/$c^{2}$) after subtracting the
combinatorial background. Corrections to the raw yields for the limited detector acceptance and
resolution, reconstruction and trigger efficiency, multiplicity effects and various analysis cuts
are determined from the full single-particle Monte Carlo simulation and analysis of the data. 

\section{Results}

\begin{figure}[tbh]
\centering
\includegraphics[width=0.49\textwidth,height=4.5cm]{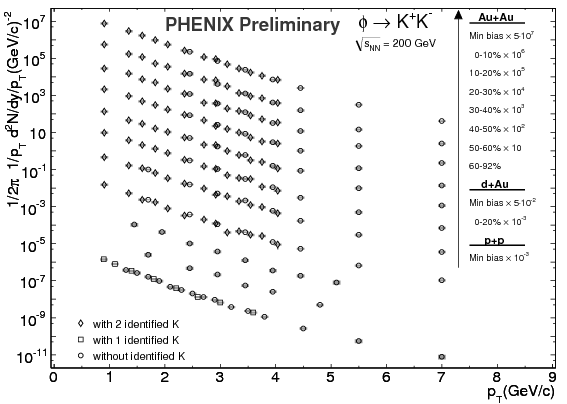}
\includegraphics[width=0.49\textwidth,height=4.5cm]{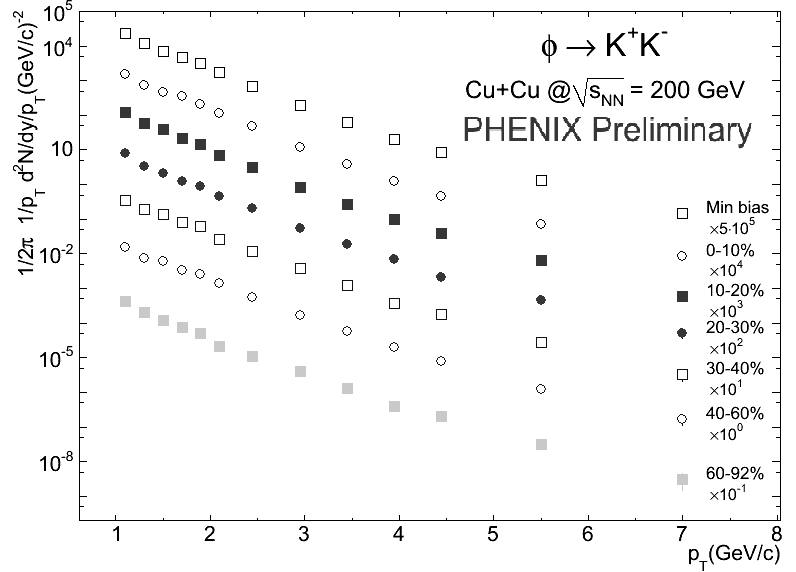}
\caption[]{Invariant $p_T$ spectra of the $\phi$-meson measured in (left) $p$+$p$,
$d$+Au, Au+Au and (right) Cu+Cu collisions at $\sqrt{s_{NN}}=200$~GeV by the PHENIX experiment.}
\label{fig:auau_cucu_spectra}
\end{figure}

\begin{figure}[tbh]
\begin{minipage}[t]{0.53\linewidth}
\centering
\includegraphics[height=5.3cm]{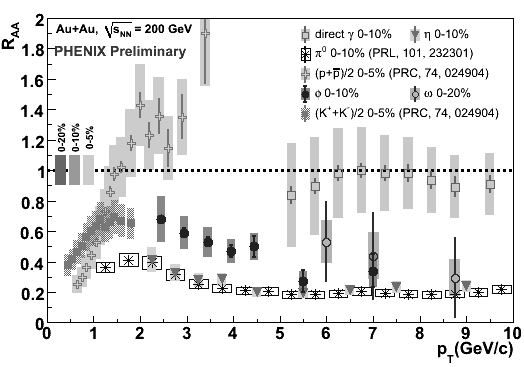}
\caption[]{Nuclear modification factor R$_{AA}$ in central Au+Au collisions
as a function of $p_T$ for $\pi^0$, ($K^+$+$K^-$)/2, $\eta$, $\omega$, (p+$\bar{\rm p}$)/2, $\phi$ and
direct $\gamma$.}
\label{fig:figure1}
\end{minipage}
\hspace{0.5cm}
\begin{minipage}[t]{0.45\linewidth}
\centering
\includegraphics[height=5.0cm]{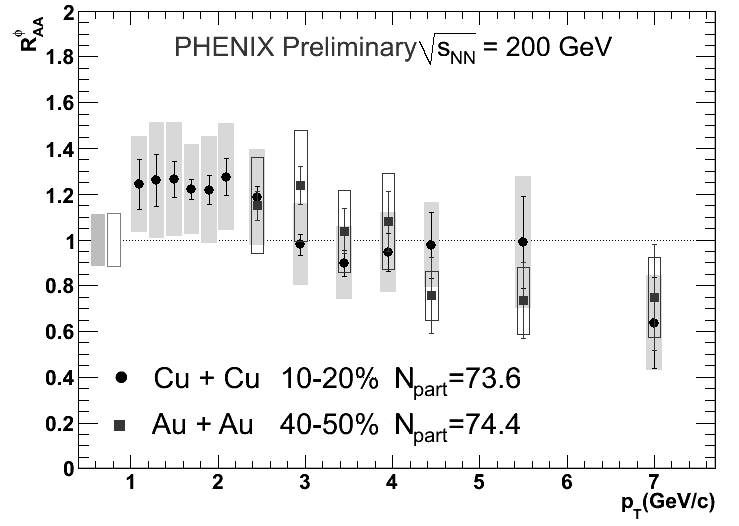}
\caption[]{Comparison of the $\phi$'s R$_{AA}$ measured in Au+Au and Cu+Cu collisions for similar
number of participating nucleons $\langle N_{part}\rangle$.}
\label{fig:figure2}
\end{minipage}
\end{figure}

The $\phi$-meson invariant $p_T$ spectra measured in $p$+$p$, $d$+Au, Cu+Cu and Au+Au collisions
are summarized in Fig. \ref{fig:auau_cucu_spectra}. 
Left panel of the figure shows good agreement between the various techniques.
Using the combined $p$+$p$ results as a reference, the nuclear modification factor has been 
derived for the $\phi$-meson in $d$+Au, Au+Au and Cu+Cu collisions. In minimum bias and most
central $d$+Au collisions, the modification factor R$_{dA}$ shows no suppression of $\phi$-mesons
\cite{vryabov}. 
Fig.~\ref{fig:figure1} shows the R$_{AA}$ for the $\phi$ in the most
central Au+Au collisions over a $p_T$ range of 2.45-7~GeV/$c$. There is an ongoing work to extend
this $p_T$ range towards low-$p_T$. The R$_{AA}$ for $\pi^0$, ($K^+$+$K^-$)/2, $\eta$, $\omega$,
(p+$\bar{\rm p}$)/2, and direct $\gamma$ are also shown in Fig.\ref{fig:figure1} for comparison.
High-$p_T$ direct photons show no suppression up to $p_T\approx$14~GeV/$c$ \cite{reygers}. Protons
are enhanced at $p_T>$1.5~GeV/$c$ \cite{ppg030}. Neutral pions and $\eta$ mesons follow the same suppression
pattern \cite{pi0_eta_suppress}. $\phi$-mesons appear to be less suppressed compared to $\pi^0$ and $\eta$ mesons in the
$p_T$ range of 2.45$<p_T<$4.5~GeV/$c$. At higher $p_T$ ($>$5~GeV/$c$) the amount of suppression of
$\phi$ and $\omega$ mesons could be similar to that of $\pi^0$ and $\eta$, unfortunately both
statistical and systematic errors are too large for a more conclusive statement. The similarity
between the suppression patterns of different mesons at high-$p_T$ supports the concept of
high-$p_T$ particle production via the fragmentation of partons outside the hot and dense medium.
It is not clear whether or not the R$_{AA}$ for the kaon (also containing a strange quark) follows
the trend of the $\phi$, since the present measurements have no overlap in $p_T$.

\begin{wrapfigure}{r}{6.7cm}
\flushright
\includegraphics[width=6.5cm]{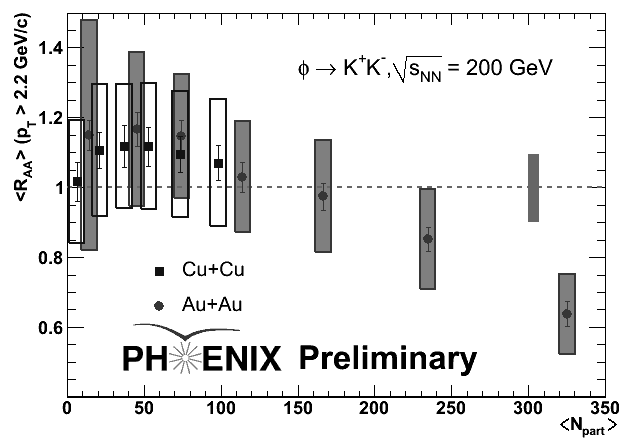}
    \caption{$\phi$-meson's integrated R$_{AA}$ for Au+Au and
	Cu+Cu collisions at $\sqrt{s_{NN}}=200$~GeV.}
	  \label{fig:intraa}
\end{wrapfigure}

Comparative studies of Au+Au and Cu+Cu data sets allow to test how sensitive is the particle
production at high-$p_T$ to the collision geometry. The $\phi$'s R$_{AA}$ measurements performed
in Au+Au and Cu+Cu collisions for similar number of participating nucleons $\langle
N_{part}\rangle$, i.e. similar energy density, yield similar results. Fig.~\ref{fig:figure2} shows
no significant difference in the suppression patterns for 40\%-50\% central Au+Au ($\langle
N_{part}\rangle$=74.4) and 10\%-20\% central Cu+Cu ($\langle N_{part}\rangle$=73.6) collisions in
the overlapping $p_T$ range. This is illustrated more generally in Fig.~\ref{fig:intraa} where the
integrated nuclear modification factor is shown as a function of $\langle N_{part}\rangle$.
Despite the differences in the amount of suppression between the $\phi$ and $\pi^0$, good agreement
in the suppression levels of the same meson is observed in Au+Au and Cu+Cu collisions for similar $\langle
N_{part}\rangle$ \cite{buesching}.

\section{Summary}
The PHENIX experiment has measured the $\phi$-meson production in $p$+$p$,
$d$+Au, Cu+Cu and Au+Au collisions at $\sqrt{s_{NN}}=200$~GeV via the K$^+$K$^-$ decay channel.
The suppression pattern for the $\phi$-meson measured in central Au+Au collisions is drastically
different at intermediate-$p_T$ from that for $\pi^0$ and $\eta$ mesons. The $\phi$'s R$_{AA}$ for
Au+Au and Cu+Cu collisions shows a similar dependence on $p_T$ and $\langle N_{part}\rangle$. 


\section*{Acknowledgments} 
The author acknowledges support by the Israel Science Foundation, the MINERVA Foundation and the  
Nella and Leon Benoziyo Center of High Energy Physics Research.

\end{document}